# The *nonlinear-by-switching* systems
# (a heuristic discussion of some basic singular systems)

*Emanuel Gluskin*

Holon Institute of Technology, 58102, Israel, *and* Electrical Engineering Department of the Ben-Gurion University, Beer-Sheva 84105, Israel.
Email: gluskin@ee.bgu.ac.il.
http://www.hit.ac.il/departments/electronics/staff/gluskin.htm

**Abstract**: Electronics has changed greatly during recent decades, and some its basic concepts should be revisited. Starting from the *sampling procedure*, we consider some mathematical, physical and engineering aspects related to *singular*, mainly switching, systems. Since the field of such systems is very rich in content, a certain line of treatment had to be chosen, making the work a theoretical introduction to the field of the systems. The focus is on the conditions for a singular system to be linear or nonlinear, and one studies more deeply what "nonlinearity" is (can be).

In order to uniformly present mathematical, physical and circuit arguments, the work is given the frame of a discussion in which the relevant specialists participate.

## 1. Introduction

Switched and sampling systems have become very common, in particular, in power electronics and in communication engineering, and a wide discussion of such, in general, singular, systems, completed by analysis of the conditions for their linearity and nonlinearity, is timely and should be useful.

We thus consider only the most basic properties of some singular systems, but from the various (circuit, mathematical and physical) points of view. Though many interesting aspects of this very wide field cannot be touched on here, the line of treatment and the finally suggested unprejudiced position regarding possible applications may hopefully be interesting to both a beginner (or a teacher who even can be a pure mathematician) and a wide-profile specialist in such systems.

Our main notations and terminology are as follows.

$t$ – time;  $t_o$ – initial moment;

$t^*$, $t_k$, k = 1,2, … -- instants of the singular operation (of sampling, switching, etc.);

$t_1$(mod $T$) – $T$-periodically repeated point (instants) of singularity;

**t*** -- the vector/set $\{t_k\}$;

**x**($t$) – vector of the state-variables $\{x_s\}$ (the unknown functions to be found); s = 1,2,…;   **x**$_o$ = **x**($t_o$);

$u(t)$ – the unit-step function (0 for $t < 0$, 1 for $t > 0$);

sign[$x$] – the "signum" function, i.e. -1 for $x < 0$ and 1 for $x > 0$; sign[$x$] = - $u(-x)$ + $u(x)$;

**u**($t$) – system inputs ("controls");

LTV – *linear time-variant*; LTI -- – *linear time-invariant*; NL – *nonlinear*. Since we always assume that a purposely switched system is indeed operated so that the switchings occur, any switched system here is either LTV or NL, and never LTI. Thus, "*LTV or NL*" is equivalent here to "*linear or nonlinear*". All that is said in this work about switched systems, is also correct regarding any system intended for *singular* operations, e.g. a sampling system.

$f(t)$ -- the function to be sampled (can be a component of **u**($t$));

$f_{ref}(t)$ – supporting (reference) input of a comparator, a known function;

$f_{trig}(t)$ – the (informational) main trigger input of a comparator, which is known (prescribed) for an LTV system and a priori unknown for an NL one;

$F(\{.\},.)$ – a *known* function of its arguments that by themselves can be some unknown parameters;

$K(t)$ – kernel of an integral, or Green's function, or the impulse(shock)-response;

$L, C, R$ -- circuits elements; $A, B$ – some positive parameters of amplitude-type;

$\vec{v}$ -- velocity (of a liquid flow) which is a function of $t$ and the spatial coordinates $x,y,z$;

$\nabla$ -- the gradient operator;

"*singularity*" – presence of a point at which sampling or switching occurs. The singularity can be expressed in a jump of any derivative (of order 0,1,2, …) of a time-function in the system under study. *Linear singular system* necessarily is an *LTV system*. As a rule, such a system is a switched or a sampling system, but passive elements with singular characteristics (appearing only at the end of the discussion) also are legitimized to constitute a singular system. In fact, in a passive singular element some internal physical "switching" also occurs.

$\alpha$, $\beta$ -- parameters relevant to the degree of nonlinearity of a system; $\alpha$ *compares* an LTV and an NL systems; $\beta$ characterizes either an LTV, or an NL system *taken by itself*.

## 2. A discussion by a Mathematician (M), a Physicist (P), and an Engineer (E) about the linearity of the sampling procedure and some associated singular systems problems

**M**: I once heard from a physicist the opinion that "our intuition is linear". However it seems that today the concepts of linearity and nonlinearity are similarly important. Engineers have learned how to formulate nonlinear problems rigorously, which is necessary if mathematical attention is to be attracted.

**P**: Rigor in science and engineering can not always be reachable. Thus, one of my students considered recently an electrical circuit with an ideal voltage source, and proved by means of computer simulation that a nice chaotic process can be obtained in it. In the laboratory he discovered that the internal impedance of the real voltage source is *unknown in the wide frequency range of the chaotic process developed in the circuit*, i.e. that the real circuit and thus the true equations are unknown to him. Fortunately, the chaos he obtained in the real circuit was also sufficiently nice, but we had some unpleasant minutes.

Nevertheless, I understand that an idealization can be necessary in order to provide the rigor needed for a mathematician to be involved.

However, how you define *linearity*?

**M**: First of all, the concept of linearity is based on the concept of *summation*. Thus, for instance, the equality

$$(f_1 + f_2)(t^*) = f_1(t^*) + f_2(t^*) \qquad (1)$$

expresses linearity of the functional (i.e. a map of a function to a number)

$$f(.) \rightarrow f(t^*) \qquad (2)$$

which I call "evaluation", and you "sampling". One can speak, of course, about the more general form

$$(af_1 + bf_2)(t^*) = af_1(t^*) + bf_2(t^*) \qquad (1a)$$

with some constants '*a*' and '*b*', also introducing what one calls "linear scaling"; $(af_1)(.) = af_1(.)$.

**P**: Is (1) a *definition* of the function $(f_1 + f_2)(.)$, or a property *to be proved* of the sampling map (2)?

**M**: Equation (1) is the *definition* of $(f_1 + f_2)(.)$. The role of summation in the axioms of linear space [1] does not allow one to interpret (1) as a feature to be proved. An equation such as (1) is correct axiomatically, and the linear scaling used in (1a) is also correct axiomatically [1]. I could speak here about any linear "operator", if you like abstract concepts.

However, consider that when one sums $f_1(t)$ and $f_2(t)$, there is, in principle, the possibility of mutual cancellation of some poles, which can make the range of definition of $(f_1 + f_2)(t)$ be wider than that of the sum $f_1(t) + f_2(t)$ where each term has to be defined independently.  In this sense $(f_1 + f_2)(.)$ is not quite the same as  $f_1(.)$ + $f_2(.)$.

**P**:  I see that the rigor approach is not quite so simple.

**M**:  Compared to the wide range of application of the concept of linearity, it is simple. However, it is now your turn, -- how you define *nonlinearity*?

**P**:   *Frightening and interesting!*  Yes, yes.  I know that "nonlinear" means "not a linear one", but this is as unconstructive as to say that "curve is not a straight line". Nonlinearity is an *initial, non-definable* concept, and, if you wish, linearity is its particular case.  I ask for your pardon for not even being ready to consider any axioms here.

**M**:  Why not to simply say that for a nonlinear system the test of linearity fails?

**P**:   This is similarly non-constructive.  You just added the word "system".  The test of linearity is formulated in terms of input-output map of the system (or operator), and *not* in terms of its physical structure.  However in any concrete problem, especially a nonlinear one, it is necessary to give the structure in detail.
   Believe me that you cannot generally define what nonlinearity is as you cannot define what beauty is.  Just forget any such idea.  However we both agree that the nonlinear problems have opened a new great world of modern science for us, and this is the real "definition".
   Let me, however, return to sampling.  I heard once from an electrical engineer the opinion that sampling is a *singular* procedure.  What does it mean?  I teach these guys only basic things, and then they bring me something not easy but always new and interesting.

**E**:  Dear Professors, this discussion interests me because the sampling procedure is important in my electronic circuits, and, in general, because *singular* (sampling, switched ...) circuits are, today, the main tool in practical electronics, as were LTI circuits at the beginning of the electronics era.  As one deals more today with quantum objects in physics and discrete models in mathematics, one also deals more today with singularities and discrete operations in electronics.  Thus, this is a good and timely topic to share mutually completing opinions, and let me contribute to the discussion.
   I do not deal with any axiomatization and do not care whether (1) is correct by definition, or has to be proved as a theorem, but one should be sure regarding (1) that he indeed performs/realizes *linear* sampling, and it is thus important for any such analysis to define the relevant *system*.  The comparison of linear and nonlinear sampling or, rather, singular (e.g. also switched) systems, is, I think, one of the most beautiful aspects of modern circuit theory, and it should influence the mathematical and physical outlooks on different singular problems.  Let us consider the realization aspect.
   You agree, of course, that in order for the map (2) be linear, i.e. for (1) or (1a) to be surely correct, the instant $t^*$ must be *fixed*, i.e. *independent of the function(s) to be sampled*.  For this to be physically provided, the *trigger*, $f_{trig}(t)$, -- i.e. an input of the subsystem that controls operation of the *sampling device* whose input is $f(t)$ and output $f(t^*)$, -- has to be generated absolutely independently of $f(t)$.

Thus, we immediately come to the thesis (see also [2,3]) saying that we have *to distinguish between linear sampling systems in which all the triggers are independent of the functions to be sampled, and nonlinear sampling systems in which at least some of the triggers are dependent on these functions.* In the latter case there will be different "$t^*$" in the different terms of (1), and then (1) can be correct only under some special conditions.

Consider these realization basics using Fig. 1.

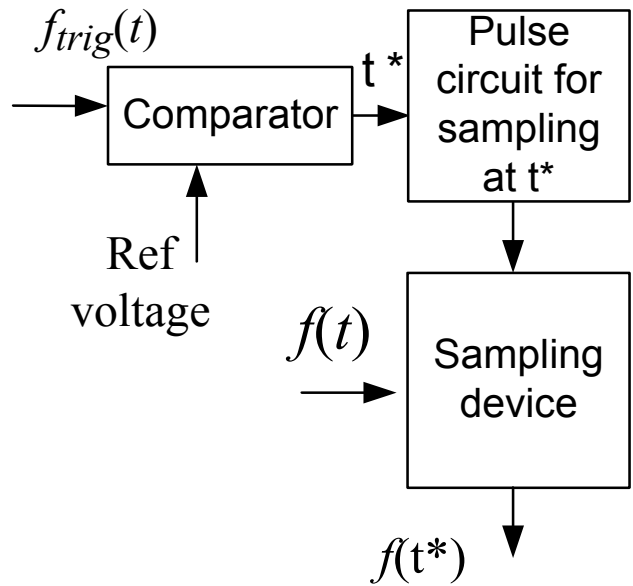

Fig. 1:   Illustration to the realization of sampling of $f(t)$ at $t^*$.   Below, this scheme is also interpreted as a subsystem included in a dynamic system. For linearity of the sampling, independence of $t^*$, and thus of $f_{trig}(t)$, of $f(t)$, is required.

Let us go from the end to the beginning of this scheme. The functions to be sampled by the sampling device are "$f(t)$". For the sampling at $t^*$, a triggering pulse has to be generated by the "Pulse circuit". The instant $t^*$ is defined by the output of the comparator that compares $f_{trig}(t)$ with some known reference voltage that, in the simplest case, may be a battery's constant voltage; $t^* = t\colon f_{trig}(t) = f_{ref}(t)$. The main point, -- absolutely not seen in (1), -- is the nature of the function $f_{trig}(t)$. In order $t^*$ be independent of $f(t)$ (the linearity), $f_{trig}(t)$ has to be independent of $f(t)$, say, specially generated by some synchronizer which is not shown in the figure but is an important part of the system.

If, however, $f_{trig}(t)$ is one of the functions "$f(t)$", or is influenced by one of the "$f(t)$", then $t^*$ depends on "$f(t)$" and the system is nonlinear. One understands that if in (1) $t^*$ is not the same everywhere, then even if (1) can be provided, one can not see in it any linear relation, or an identity.

You can see now the importance of the realization aspects associated with the "innocent" $t^*$ in (1). The argument $t^*$ which represents (just by its discrete nature) the *singularity of a system*, is responsible for the *linearity or nonlinearity* of this system. One can even consider here that a connection between $f_{trig}(t)$ and $f(t)$ may arise not as planned, but because of electromagnetic (radiation) interferences in the circuit. Such interference is a matter of physics, but the very fact that *nonlinearity* can arise because of a weak and per se linear electromagnetic coupling stresses the importance of the general understanding of what nonlinearity can be.

**P**: Your comment on electromagnetic coupling reminds me of the good old days when I was a young student working in a laboratory and having problems with the synchronization of a simple oscilloscope. Then, we were not seeking any "chaos" in the non-synchronized state of everything jumping on the screen, but were trying to provide the conditions for synchronization that, in your terms, is prescribing, by means of a synchronizer, such parameters as $t^*$.

**E**: Not every nonlinearity of the type we discuss causes chaos or non-synchronization, but if you want to *surely* prevent any such trouble, -- you do have to provide the linearity of the basic supporting processes in the oscilloscope, in terms of prescribed "$t^*$".

**P**: I also see a specific but interesting point in the use of a comparator for the definition of $t^*$. While from the pure mathematical positions one can assume that in (1) the functions are complex-valued, your Fig. 1 suggests that we should speak here only about real-valued functions. Indeed, the necessary physical comparison ('<', or '>') of the $f_{trig}(t)$ with the threshold level $f_{ref}(t)$ in the comparator relates only to real-valued variables. Complex values cannot be thus compared. Generally, comparators avoid complex considerations.

It becomes clearer why the $\delta$-function is always defined in physics and engineering as real-valued. This is because it is inherently associated with sampling. Indeed, its main action is

$$\int_{-\infty}^{\infty} f(t)\delta(t-t^*)dt = f(t^*) , \qquad (3)$$

which may be interpreted in terms of the blocks of Fig. 1. Since $f_{trig}(t)$ must be real valued, the possible connection of $f_{trig}$ with "$f$" (I use here the option of the nonlinearity) forces "$f$" in (1) and (2) to be real valued, and then also the $\delta$-function in (3) has to be real-valued.

**E**: For me all this is absolutely clear. I consider sampling to be a part of a *measurement*, but measurements always relate to real values. Thus, $\delta(t)$ should be real. I do not know who at all needs complex $\delta$-function.

**M**: I guess, *any mathematician*. Sorry, but I simply see in your whole scheme just a *specific realization* of a mathematical object, here map (2), while *for me the $\delta$-function is the very map* (2) *and nothing else*. Not having to be concerned with the topic of realizations, I can see the functions to be sampled as complex. I know that physicists like to define the $\delta$-function as the limit of the sequence of some real-valued sharp pulses. However also this is just a realization of the evaluation (2). Are you sure that for a complex $f(.)$, the $\delta$-function in (3) *must* be real-valued ?

However let me try to formalize what we have regarding Fig. 1. We were initially focused on the description of the functions "$f(.)$". You suggest that we should focus on the specific degree of freedom associated with the *choice* of $t^*$, and insist that the situation regarding realization of the sampling at $t^*$ is sufficiently important for a mathematician to keep it in mind. However, theoretically, the concept of linear sampling is very simple; you fix $t^*$ by taking $f_{trig}(t)$ from an independent stable synchronizer, and thus ensure the linearity. Your basic point regarding "linear-or-nonlinear" in the realization terms is clear, but I would better understand your interest in the realization of $t^*$ if you would also consider nonlinear systems.

The idea of obtaining nonlinearity by means of control of $f_{trig}(t)$, i.e. directly in terms of time functions involved, seems to be elegant, but your point should be presented more constructively.

**E**: Of course I intended to also show you the mathematical/*analytical* aspects of the topic of realization.

The possibilities of $f_{trig}(t)$ (or $t^*$) be, or not, connected with $f(t)$, suggests that one should consider the problem of the linearity of (1), or, rather, of any singular system, in terms of *state-variables* [4-8] that in electronics are some initially unknown currents and voltages generated in a circuit, and in the other disciplines are also some well-measured parameters. Thus the structural aspect is inevitably introduced.

In what follows, parameters like $t^*$, are not necessarily *sampling* instants. It is most suitable to speak here about *switching* operations, when some linear elements replace each other (are "changed") at the switching instants. The criteria for linearity or nonlinearity are, finally, the same for any such singular system.

Let us compare the obviously linear (LTV) system [4-7]

$$d\mathbf{x}/dt = [A(t)]\mathbf{x} + [B(t)]\mathbf{u} \qquad (4)$$

to the obviously nonlinear

$$d\mathbf{x}/dt = [A(\mathbf{x},t)]\mathbf{x} + [B(\mathbf{x},t)]\mathbf{u} \ . \qquad (5)$$

Here **x** is vector of the state-variables, and **u** is vector of the inputs that, in the context of sampling (not necessary below), include the functions "$f(t)$". Vectors **x** and **u** are thus connected through the dynamics of the whole system.

Actually, I am going to repeat what I already said re Fig. 1 about linearity and nonlinearity, but in the constructive terms permitted by (4) and (5). The role of the functions $f_{trig}(t)$ is now seen via dependence of the matrixes [A] and [B] on $t$ and/or **x**.

At each switching moment $t_k \in \mathbf{t}^*$ some parameters of the system are changed, and thus [A] and [B] depend on $\mathbf{t}^*$ that by itself is defined by $f_{trig}(t)$. Thus, if $f_{trig}(t)$ is connected with **x**, then $\mathbf{t}^*$ and the matrices depend on **x**, and the system is nonlinear. The consideration of the dependence of $\mathbf{t}^*$ on **x** includes, in particular, the sampling problem because **x** is connected with **u** which in such a problem includes $f$.

It becomes clear that the general role of the instants of the singularity can be formalized as follows.

Since switchings change, instantaneously in time, the parameters of the system, we first symbolically write for *any* switched system

$$[A] = [A(\mathbf{t}^*(.), t)] \quad \text{and} \quad [B] = [B(\mathbf{t}^*(.), t)] \qquad (6)$$

and then consider, in terms of $\mathbf{t}^*(.)$, whether this system is linear or nonlinear.

In the linear case, set/vector $\mathbf{t}^*$ is *known* a priori:

$$\mathbf{t}^*(.) = \mathbf{t}^*(t) \ , \qquad (7)$$

i.e. the switching instants are prescribed, as it is assumed regarding *t\** in (1), and [A] and [B] are prescribed/known as time functions.

In the nonlinear case,

$$\mathbf{t^*}(.) = \mathbf{t^*}(\mathbf{x}) \qquad (\text{or } \mathbf{t^*}(\mathbf{x},t)), \tag{8}$$

i.e. **t\*** is *unknown* a priori and depends on the state-variables to be found. For (8) to be provided, at least one of the "$f_{trig}(t)$" *has to be one of the $x_s$, or depend on such a function*. Recall that in terms of Fig. 1 and eq. (1), this means that *t\** depends on the function being sampled.

Thus, respectively to (7) and (8), we obtain, via (6), equations (4) and (5).

Let us denote a singular system as a "**t\***-system", or a **t\***(.)-system, its linear version as a **t\***(*t*)-system, and its nonlinear version as a **t\***(**x**)-system. Then, Fig. 2 schematically illustrates a very general outlook on singular systems composed of inherently linear elements.

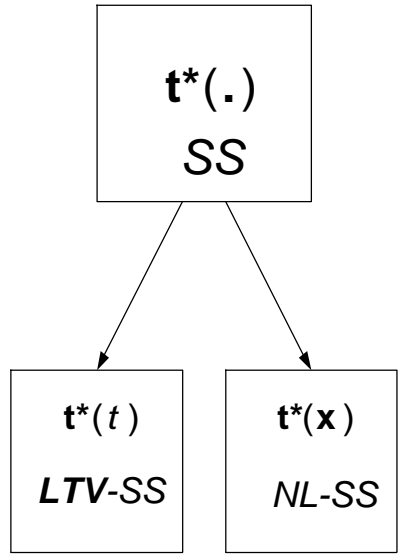

Fig. 2: The schematic classification of singular systems (SS) as linear time-variant (LTV) and nonlinear (NL), based on the principle of the points of singularity being or not being dependent on the unknown state variables to be found. Below, passive elements having singular characteristics (i.e. some internal physical switching) are also argued to be relevant to this scheme.

Observe that the LTV and NL versions of an SS are *definitionally close* because the *singularity* of switching in both versions is the same, and the terms "LTV" and/or "NL" cannot be separated (detached) from the term "SS" in this figure. This is very different from defining nonlinearity of a nonsingular system via the details of the analytical "characteristics" of some *per se* independent elements included in this system. The term "nonlinear switched systems" is even not so good for the point. I would suggest: "***nonlinear-by-switching** systems*".

Anyway, if we write, at the "first stage" of the analysis, according to (6)

$$d\mathbf{x}/dt = [A(\mathbf{t^*}(.), t)]\,\mathbf{x} + [B(\mathbf{t^*}(.), t)]\mathbf{u}, \tag{9}$$

then, in principle, we have to add to this equation a description of the conditions defining **t***(.). For the nonlinear case, such a condition can be schematically written as (here, '→' means "*influences*"):

$$\exists x_{s'}(t) \to f_{trig\ m} \to [f_{trig\ m}(t) = f_{ref\ m}(t)] \to t_{k'}$$

meaning that at least one of the state-variables, $x_{s'}$, influences the informational input of one of the triggers, number *m*, and thus also the certain switching instant $t_{k'}$ that is the solution of the equation in the brockets.

A more general, and this time a "closed" scheme for the shifting *nonlinearity*, relevant to the dynamic context and symbolically describing the development of the nonlinear process **x**, is

$$\mathbf{x} \to f_{trig} \to \mathbf{t^*} \to \text{the structure of the system (i.e. [A] and [B])} \to \mathbf{x}. \quad (10)$$

**M**: How do you actually introduce **t*** into [A] and [B]?

**E**: All the parameters included in Kirchhoff's equations are seen as instantaneous time functions. Take first [A(*R*)] for a constant resistance *R* in an LTI system and then pass on to a switched system by setting

$$R \to R(t) = R_1 u(t_1-t) + R_2 u(t-t_1) ,$$

where *u*(*x*) is the unit-step function that equals 0 for *x* < 0 and 1 for *x* > 0. That is, $R(t) = R_1$ for $t < t_1$, and $R_2$ for $t > t_1$. Parameters of the types *R*, *C* and *L*, and also parameters characterizing dependent sources, appearing in the matrices, can be thus made singularly time-dependent by means of switchings. Obviously, some [A($t_1,t$)] is obtained.

**M**: However, [A(**t***(**x**), *t*)] and [A(**t***(*t*), *t*)] look here similar. Can one "feel", at the stage of the solution of (9), i.e. in terms of **t***(.) and not yet considering (10), that the system is nonlinear?

**E**: The symbolism of the notations **t***(*t*), and **t***(**x**) has to be stressed. The instants of singularity here are *certain numerical values*, and, for example, for the integration by *t* of a function of *t* and **t*** = {$t_k$}, one need not know what are {$t_k$(.)}; they remain as some parameters for further analysis of the problem.

If the inputs **u** are given, then at this stage one indeed do not feel the nonlinearity, and the solving processes for NL and an LTV systems are similar. However, the nonlinearity is immediately felt if one makes a test of linearity, e.g. the scaling one, **u** → *k***u**, with a constant *k*. Then, contrary to the case of **t***(*t*), some respective changes occurring in **x** and dependent on *k* cause **t***(**x**) to be moved, i.e. [A(**t***(**x**), *t*)] is changed, and one sees from (9), or (5), that **x** does *not* become just *k***x**, which means nonlinearity.

Later, I shall give an example of a nonlinear system, demonstrating the role of {$t_k$}.

**M**: Dynamic systems are very important, but your notation "**t***(.)-system" relates to any singular system and it is independent of the state-space outlook. What would be bad in using instead of (9) the purely algebraic, constructive form

$$\mathbf{x} = [D(\mathbf{t^*}(.), t)]\mathbf{u}(t), \qquad (11)$$

with a matrix [D] (whose physical units are those of [B], for each element, multiplied by *second*), having both of the possibilities of $\mathbf{t^*}(\mathbf{x})$ and $\mathbf{t^*}(t)$?

**E**: Including this case, a symbolic diagram of the possible influence of $f(t)$ on $\mathbf{t^*}$ is as shown in Fig. 3.

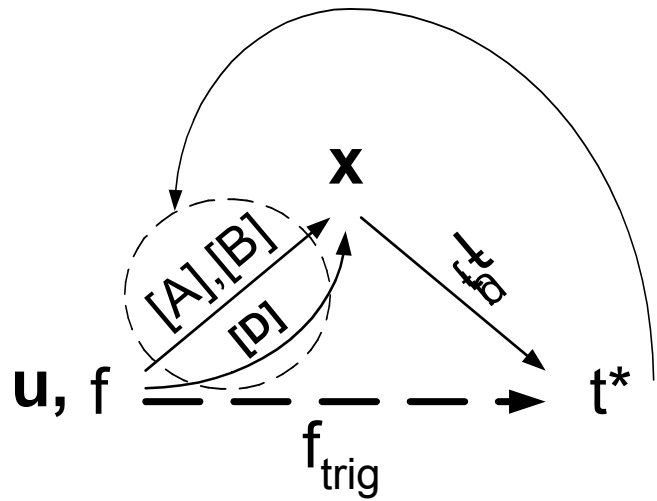

Fig. 3: The shifting *nonlinearity* i.e. the influence of $f(t)$ on $\mathbf{t^*}$ from the realization point of view. The realization is either done by using [B] and [A], i.e. solving differential equations, or using [D], i.e. purely algebraically as for equation (11). The feedback to the encircled symbols means influence of the switching instances on the *structure* of the system. This is the essence of the nonlinearity as it is expressed by (5), or by (11).

**M**: You write your nonlinear system (5) unusually. One would write such a system as

$$d\mathbf{x}/dt = \Phi(\mathbf{x},t) \qquad (12)$$

where vector-function $\Phi$ has the proper dimension. I understand that your writing (12) in the form (5) is intended to allow one to *compare* the nonlinear and linear "versions" (4) and (5) of a singular system, which is your very point. You wish to see the "structure" of such linear and nonlinear systems *in the same terms*, i.e. via the elements influencing the matrices, and your equation (9) even unites (4) and (5). The comparison of the linear equation (4) with (12) is not so immediate as the comparison with (5). The usual transfer from (12) to (4) is associated with *linearizations* of a nonlinear system, which is irrelevant to the comparison with (5). Equation (4) is not a linearization of (5), but its particular case. I hope that I understand this general point correctly.

**E**: Yes, you formulated the point precisely, and despite its formal simplicity, it was not easy to come to it. The decision to replace (12) by (5) is a result of a long-term research work. Though (12) seems to be more general, (5) has, so to speak, some optimal degree of generality, more suitable for understanding the situation regarding the system *realization*. Let me define this position as a mathematical realism.

Form (5) can be advantageous in defining and analyzing a structure, in particular, when one can think in terms of subsystems, and when the possible nonlinearity of the whole system is defined by mutual connections of the subsystems. I believe that in the sequel of our discussion such an example will naturally appear.

**P**: If I consider only the point of nonlinearity, i.e. the dependence of the structure on **x**, without the context of switching, and try to see the general heuristic advantages of writing (5) instead of (12), then the dynamic interpretation and equation (5) makes me recall the basic hydrodynamics equation. Let us compare the nonlinear "system"-term of (5)

$$[A(\mathbf{x}, t)] \mathbf{x}$$

with the nonlinear physical term

$$(\vec{v}\nabla)\vec{v}$$

appearing in the full time-derivative $\frac{d}{dt} = \frac{\partial}{\partial t} + (\vec{v}\nabla)$ of the velocity $\vec{v}$ in the Navier-Stokes equation [9,10]

$$\frac{\partial \vec{v}}{\partial t} + (\vec{v}\nabla)\vec{v} = \nu\Delta\vec{v} \ , \qquad (13)$$

where $\nu$ is the viscosity of the liquid.

*In both cases the structure of the system expressed in the mathematical operators acting on the unknown variables/fields is given in terms of these variables/fields.*

Assume that one does not know anything about the Navier-Stokes equation, but when observing a liquid flow thinks in terms of system-structure and sees the velocity vector field as both the variable to be found and as the "structure" of the liquid system. Then, one can readily understand, in view of (5), that the hydrodynamic equations intended to describe the flow must be nonlinear. Then, turbulence can be naturally accepted by him as a "chaos" generated in the nonlinear "system", while the case of some, on the whole laminar, prescribed in some way flow carrying relatively rare vortices will be understood as a linearization of the nonlinear equations.

**M**. I feel that this is of some heuristic interest, but can this be helpful in modeling (calculating) of a hydrodynamic flow?

**P**: I do not know, but analytical solution of the Navier-Stokes equation is considered to be impossible in the general case, and any advance would be important here.

**E**: It is easy to rewrite the vector equation (13) in the proper matrix form. Just move the nonlinear term to the right-hand side and multiply this side by the *unit matrix* [*I*], 3×3,

$$\frac{\partial \vec{v}}{\partial t} = [I](-\vec{v}\nabla + \nu\Delta)\vec{v} \ ,$$

thus obtaining

$$\frac{\partial \vec{v}}{\partial t} = [A(\vec{v})]\vec{v} + [B]\vec{v} \qquad (14)$$

where

$$[A(\vec{v})] = -[I](\vec{v}\nabla) = \begin{pmatrix} -\vec{v}\nabla & 0 & 0 \\ 0 & -\vec{v}\nabla & 0 \\ 0 & 0 & -\vec{v}\nabla \end{pmatrix}, \quad \vec{v}\nabla \equiv v_x \frac{\partial}{\partial x} + v_y \frac{\partial}{\partial y} + v_z \frac{\partial}{\partial z},$$

and

$$[B] = \nu[I]\Delta \equiv \nu \begin{pmatrix} \nabla & 0 & 0 \\ 0 & \nabla & 0 \\ 0 & 0 & \nabla \end{pmatrix}, \quad \Delta = \frac{\partial^2}{\partial x^2} + \frac{\partial^2}{\partial y^2} + \frac{\partial^2}{\partial z^2}.$$

The system approach to the propagation of a perturbation of a liquid flow suggested in [10] is of some relevance here. The perturbation is considered there as the "input" of a "system" which is carried by a laminar background flow. This means a linearization of (13) or (14), which allows in [10] the concept of transfer function to be applied, with the Laplace transform done by *t*. In such an input-output problem, with the artificially fixed/defined boundaries (the placements of the input and the output), the partial derivative by time in (14) has, finally, the same meaning as the full derivative in (5) or (9). However, whether or not this formulation in terms of the matrices-operators really helps is a question to many mathematicians.

**P**: I also have a more pragmatic comment and a question. Switched electronics circuits include many modern chaotic circuits, some of which (e.g. [11] and references there) are used in our students' physics laboratories. Since any chaotic circuit is nonlinear, your classification of switched circuits as linear or nonlinear should be of some interest for physicists. However, the nonlinearity that may actually occur here is only of the "shifting"-type $F(t-t^*)$, or $F(\{t-t_k\})$, where $F(.)$ is a known function (waveform), while $t^*$, or $\mathbf{t}^* = \{t_k\}$, is dependent on some unknown functions to be found. I understand that together with the dramatically increasing importance of the switching and other singular operations in electronics, this kind of nonlinearity becomes more and more important, but I must warn you that for many this is a very unusual, specific nonlinearity. For my students such "*characteristic*", as, say, $y(x) = ax + bx^3$, is a typical example of nonlinearity. It is easy to see whether or not the latter nonlinearity is weak ($bx^3 \ll ax$), while for the case of the nonlinear shifting I cannot immediately see how to define "weak nonlinearity". This seems to be a methodological or pedagogical problem, because the concepts of "weak nonlinearity" and "strong nonlinearity" are widely used.

I am afraid that it will be not easy for me to explain all this to my students.

**E**: Formally, in the switched systems, the strong or weak nonlinearity can be defined very simply. Let me return to the example of $[A(R(t-t_1)]$ where $R(t_1-t) = R_1 u(t_1-t) + R_2 u(t-t_1)$ sometimes equals $R_1$ and sometimes $R_2$. It is rather obvious that the ratio

$$\beta = \frac{|R_2 - R_1|}{R_1}$$

can be taken as a criterion of the nonlinearity. However we know that the switching need not constitute nonlinearity; if $t_1$ is prescribed, the same ratio characterizes a *weakly- or strongly-varying linear* system. Since the ratio of the type $\beta$ can be the same for an NL and an LTV versions of a system, one has to face the fact that this ratio is *not* a sufficient criterion for the point of nonlinearity.

The introduction of nonlinearity is traditionally done by means of a "characteristic", and I see that the pedagogical problem is that one gives a magic force to this term. If a system includes an element with a nonlinear "characteristic", it is nonlinear, and one would thus create an NL system, if required. Whether this characteristic is analytical, or piece-wise-linear with some inflection point(s) which is typical for switching nonlinearity, is not so essential here. *The real point is whether or not any such characteristic at all exists*.

Any "characteristic" is defined in terms of the "inherent" state-variables, and not time. Consider, starting from the origin of the relevant axes, only one of the parallel lines shown in Fig. 4. Then we indeed have a certain "broken" $x_1$-$x_2$ line as a *certain characteristic* of an element in a system, and this is an *unchangeable feature* of the NL system, fixed for any process in the system. We just have to assume that the range of the process is such that the inflection point is indeed involved, otherwise we have an LTI system with constant parameters.

However, for an LTV-system, in which the switchings are done at *prescribed instances*, the values of the coordinate variables $x_1$ and $x_2$, of the point of singularity will be changed, i.e. the inflection point of the type 'a' will be generally *different* for the different switchings (the complete Fig. 4), and, thus, in fact, *no certain "characteristic" (that could be seen as nonlinear) exists*. Let us think, for simplicity, about a periodic process that defines the intercept **ab**, and the "family" of the curves.

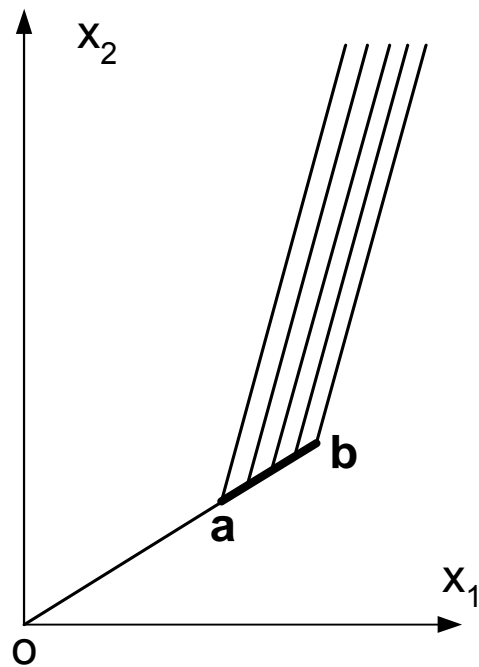

Fig. 4: The illustration of an LTV SS: when **t*** = **t***($t$) there is no certain "characteristic" of a switched unit in terms of the state-variables because the inflection point is not the same point ($x_1$,$x_2$) at each switching instant. For an NL SS, we indeed would be given, by means

of **t\*** = **t\***(**x**), a fixed inflection point in the ($x_1,x_2$)-plane (|**ab**| = 0), and a certain piecewise-linear, *nonlinear*, characteristic would exist.

Observe that this figure demonstrates two points. Firstly, a "family" of some curves in the plane of some state-variables can exist, in which case switching does not mean nonlinearity (and no "characteristic" exists), *for however strong inflection*. Secondly, one sees how the "weak linearity", *in the sense of closeness of an LTV system to an NL system*, can be defined; namely, the parameter $\alpha$ = |**ab**|/|**oa**| as a criterion for this closeness (note, -- disregarding however strong the inflection is!) is naturally introduced.

This consideration brings the possibility of defining for SS the concept of *weak linearity* (or strong nonlinearity) of an SS, -- *in the sense of the comparison of the LTV and NL versions*, -- which in terms of Fig. 4 simply means **ab** << **oa**. Thus, both the ratio $\alpha$ = |**ab**|/|**oa**| and the ratio of the type $\beta$ are important characteristics of the system. Only if $\alpha$ = 0, does such a parameter as $\beta$ characterize the degree of nonlinearity of the system.

I beg you pardon for this excursion to the prosaic world of the "characteristics", but you mentioned your students and I have tried to answer the question about degree of nonlinearity pedagogically satisfactorily.

**P**: These "prosaic" details are interesting not only as regards the introduction of $\alpha$. Can the intercept **ab** of the "family" of curves in Fig. 4 be *directly* associated with the property of *superposition* of the relevant LTV (and not any NL) system?

Schrödinger's equation is also LTV, just *spatially*–variant; thus, -- is it possible to generalize the uncertainly principle of quantum mechanics, formulated in the "state-variables" space-impulse, to a wider class of the LTV-type equations?

However, I have too many questions! Sorry!

**M**. Let me return to the main line and the equational side, considering again the role of **t\***.

Being weak or strong, the "shifting nonlinearity" is specific in the sense that it can be *united with linear operations*, because if a linear operator acts on a function of the type or $F(\{t-t_k\})$ we can obtain a function of the same type. Consider, $F_1 \to F_2$:

$$F_2(\{t_k\},t) = \hat{L}_t(F_1(\{t_k\},t)) \qquad (15)$$

where $\hat{L}_t$ is a linear operator acting on *t* (say, an integration with a kernel).

Because of the discrete nature of $\{t_k\}$, for any $\{t_k(.)\}$, the performance of the linear operation and investigation of $F_2$ as linear or nonlinear will be separated. Since theory of linear operators is very well developed [7,8], this kind of nonlinearity should have a methodological advantage.

I still do not know, however, what nonlinear effects one can obtain using this nonlinearity.

**P**: According to the previously mentioned old memories about the poorly synchronizable oscilloscope, -- *any* nonlinear effects. However, regarding the possibility of approaching $\{t_k\}$ as time constants in (15), I would like to see the circuit example promised to us.

**E**: At least for a periodic problem [12-16] which I meant, the comment on the use of linear operators is correct. In this example, a function to be found is explicitly represented by means of *its own* zerocrossings.

Consider the following circuit in Fig. 5 which includes a fluorescent lamp in a steady-state operation directly from the voltage line, i.e. at some low-frequencies.

Because of the very strong nonlinearity of the lamp's voltage-current characteristic, one can effectively use for it (as a "first approximation" [13,14]) a singular model, and the circuit equation is

$$L\frac{di}{dt} + A\text{sign}[i(t)] + \frac{1}{C}\int i(t)\,dt = B\sin\omega t \qquad (16)$$

where *L*, *A* and *C* are positive constants, and sign[*x*] = 1 for *x* > 0, and –1 for *x* < 0.

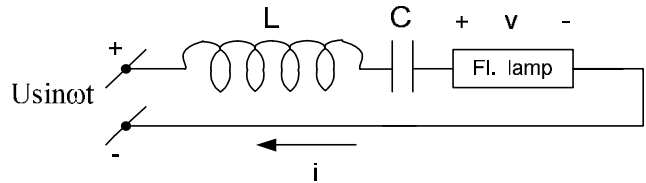

Fig. 5: The electrical circuit for equation (16). At the regular line frequencies, the lamp's *v-i* characteristic can be approximated as *v* = *A* sign *i*, or, rather (since we deal, in general, with time functions) in the "parametric form" as *v*(*t*) = *A*sign *i*(*t*), with a *zerocrossing* function *i*(*t*), so that *v*(*t*) is a *square wave* of height *A*, with a zero average, having its -/+ jumps at the -/+ zerocrossings of *i*(*t*). Apart from the singular nonlinear resistor representing the lamp, there is some small linear resistance of the inductor, not shown in this figure, which provides the needed damping properties of the kernel function *K*(*t*) used in eq' (21). This kernel function is the impulse response of the linear LC "ballast".

**E**: Because of the function 'signum', (16) is a nonlinear equation obviously.

Though being singular, the function *i*(*t*) must be continuous since otherwise d*i*/d*t* introduces into (16) an *uncompensated* δ-function. It is not difficult to show [12,13] that for $\omega = 2\pi/T \neq \omega_o = (LC)^{-1/2}$ and |*B*|/*A* sufficiently large, *i*(*t*) is a *T*-periodic *zerocrossing* time function having (as it does for sin ω*t*) two zerocrossings per period, distanced by *T*/2. For such |*B*|/*A*, the nonlinear term in (16) is the *square wave*

$$A\text{sign}[i(t)] = \frac{4A}{\pi}\sum_{1,3,5,\ldots}\frac{\sin n\omega(t-t_1)}{n}, \qquad (17)$$

where $t_1(\text{mod}T)$ is the −/+ zerocrossing of *i(t)*.

However, if we substitute (17) into (16), then in the obtained equation

$$L\frac{di}{dt} + \frac{1}{C}\int i(t)\,dt =$$
$$= B\sin\omega t - \frac{4A}{\pi}\sum_{1,3,5,\ldots}\frac{\sin n\omega(t-t_1)}{n} \qquad (18)$$

the nonlinearity that was so obvious in (16), is not at all obvious.  If $t_1$ were to be an a priori given parameter, then (18) would be a simple LTI equation.

The point is, of course, that $t_1$ belongs to the *unknown function* $i(t)$ (a state-variable) that has to be determined, and *thus* the sum (17) is a nonlinear expression, and (18) is a nonlinear equation.

Using Fourier series [12-16], or turning (18) into an integral equation, one easily obtains a function of the type (15),

$$i(t) = F(t, t_1) = Bg(t) - AF_2(t - t_1) \qquad (19)$$

with *known functions* $g(.)$, $F_2(.)$ and thus $F(.,.)$.  Of course, that $i(t)$ is represented as an explicit function of *its own* zerocrossing(s) is a case of the shifting nonlinearity.

The still unknown $t_1$ is then found from the equation that originally defines $t_1$, $i(t_1) = 0$, but now already has the *constructive form*

$$F(t_1, t_1) = Bg(t_1) - AF_2(0) = 0 ,$$

i.e.

$$g(t_1) = \frac{A}{B} F_2(0) . \qquad (20)$$

The possible necessity to determinate the zerocrossings *at a late stage of the solution* is one of the main nuances of working with **t\*(x)**.  In fact, this point has already been touched on in our discussion several times.

**P**:  Can you more comment on (19) as a case of (15)?

**E**:  In the sense of a periodic solution, (16) or (18) is equivalent to the following integral equation

$$i(t) = Bg(t) - A \int_{-\infty}^{t} K(t - \lambda) sign[i(\lambda)] d\lambda \qquad (21)$$

where $g(t)$ is a known zerocrossing function, $K(t)$ is a kernel or shock-response *of the linear LC "ballast"* (see Fig. 5), and the whole right-hand side of (19) or (21) is some $F(t, t_1)$.

**M**:  You cannot miss the aspect of "structural stability" of the waveform of $i(t)$, i.e. the requirement it to be, as $g(t)$, a zerocrossing waveform.  Consider (19) or (21).

**E**:  The circuit inductance $L$ defining the features of $K(t)$ at high frequencies causes the integral in (21) to have a bounded derivative, which is, of course, just the well-known feature of the inductor of preventing jumps of the current passing through it. Thus, for $A/B$ sufficiently small $i(t)$ has the same density of the zerocrossings as $g(t)$. This is graphically obvious when considering the zerocrossing waveform of $g(t)$ in the context of (21), but one can find the rigorous details in [12-16], especially in [12].

**M**:  I see from (20) that if

$$F_2(0) = 0 \qquad (22)$$

then $t_1$ is also a zero of $g(t)$. Thus, for a proper $K(t)$, with increase in $A$/B, starting from zero up to a critical value, the zerocrossings of $i(t)$ remain *precisely* those of $g(t)$. That is, the waveform of the non-sinusoidal of $i(t)$ is changed, but the zeros are unmoved/unchanged. The dependence of $i(t)$ on $A$ or $B$ in (19) is then linear, or, rather, *affine*, i.e. as $y(x) = ax + b$, *for* $b \neq 0$.

**E**:  In [12-16] this possibility to have $t_1$ unchanged when $B$ changes, is named *constancy* of $t_1$, and this is an important item in the theory of these circuits. As is explained in [12-16] (and one can understand this from the above equations) the constancy of $t_1$ causes the *lamp's power* $p(t) = v(t)v(t)$ to be relatively weakly dependent on the amplitude of the line voltage. Consider that the averaged consumed power $P = <p(t)>$ is the main parameter of the lamp circuit, and that on the world scale fluorescent lamps consume about 20% of all the electrical power generated. The power features of the lamp circuits, including the sensitivity of $P$ to the amplitude of the line voltage are very important.

**P**:  These energy-conversion circuits are also interesting in the sense of the quantum-physics processes in it. Adding also this to what you said, -- I think that, today, Michael Faraday would not give lectures on the chemical history of the candle, but, rather, on the quantum processes in the fluorescent lamp [17] and the associated unusual features of the lamp's *circuit*.
   Well, I see now how the focusing on the zerocrossings can lead one to constructive conclusions. However, regarding your wish not to deal with "characteristics", but only with time functions, this example is marginal. Before you write, *for the T-periodic time-functions*, the $v(t)$-$i(t)$ relation $v(t) = A\text{sign } i(t)$, you use the *characteristic* $v(i) = A\text{sign}[i]$, given in the $v$-$i$ plane, and not in the time domain.

**E**:  A realistic correction to the idealized hardlimiter model is [13,16] that the $v(t)$-$i(t)$ dependence becomes

$$v(t) = A\text{sign } i(t) + L'\, di/dt , \qquad (23)$$

where $L'$ is a small inductance of the lamp *by itself*, and one has to directly speak about a $v(t)$-$i(t)$ "relation" (or, "connection"), and not the $v(i)$ "characteristic".
   A necessary note is that in (23) '$A$' has to be taken somewhat larger than in (16). The formula for this increase in $A$ is derived in [16].
   If you still have a problem with the use of passive elements with singular characteristics in the scope of the equations (4,5), then it may interest you to see in [3] a discussion of a switched circuit with ideal diode, having the possibilities of the NL and LTV.
   We can include singularly-nonlinear *passive* elements in the consideration, and thus, in general, the singular circuits here can be: switched, sampling, and including passive elements with singular characteristics. This is a generalizing correction, to, e.g., Fig. 2.

**M**:  Sorry, but all this has become too far from me, and let me return to the main equations, collecting what we have to this moment on the main line.
   The parameters $\{t_k\}$ are just some numerical values which -- when not prescribed -- have to be finally found from some algebraic equations for $\{t_k(\mathbf{x})\}$. Strongly simplifying any such symbolic diagram as (10), one can write such an algebraic equation as $x_{s'}(t_{k'}) = 0$, for some certain $s'$ and $k'$, which, similar to (20) (where $x_{s'}$ is $i$), i.e. similarly to

$$g(t_1) - \frac{A}{B} \int_{-\infty}^{t_1} K(t_1 - \lambda) sign[i(\lambda)] d\lambda = 0 ,$$

defines $t_k(x_{s'})$ ( here $t_1(i(.))$ ).

In your general scheme, one has first of all to solve the "preliminary" matrix equation (9), $d\mathbf{x}/dt = [A(\mathbf{t^*}(.), t)] \mathbf{x} + [B(\mathbf{t^*}(.), t)]$, as is usual for LTV circuits. At this "dynamic stage" one cannot see the nonlinearity of the shifts. The parameters $\{t_k(.)\}$ remain for the final algebraic stage, and only then a solution $\mathbf{x}(\mathbf{x_o}, t_o, \mathbf{t^*}, t)$ is finally found.

I do not see any problem with this scheme, *but LTV equations are quite difficult to analyze*, and just in order to make one's analysis easier, I would suggest starting with a relatively simple and anyway necessary problem.

Consider:

$$d\mathbf{x}/dt = [A(\mathbf{t^*}(.), t)] \mathbf{x}$$

and assume that the system is LTV, i.e. $d\mathbf{x}/dt = [A(t)]\mathbf{x}$. The *transition matrix*, $[\Phi(t, t_o)]$ is then found

$$\mathbf{x}(t) = [\Phi(t, t_o)]\mathbf{x_o}$$

by a not quite easy iterative procedure [6].

For an LTI system, we have [A] constant (while $\mathbf{t^*}$ stops to have any meaning), and for $d\mathbf{x}/dt = [A]\mathbf{x}$, we have $[\Phi(t, t_o)] = [\Phi(t - t_o)] = \exp\{[A]\cdot(t-t_o)\}$. The latter expression is obtained from the matrix $\exp \int_{t_0}^{t} [A(t)]dt$, *for* [A] *constant*. However, if [A] indeed depends on $t$ (and $\mathbf{t^*}$ arises), then for the above simple exponent with the integral to be $[\Phi(t, t_o)]$ it is necessary [6,7] for [A(t)] to *commute* with $\int_{t_0}^{t} [A(t)]dt$. As the point, -- it would be interesting to clarify how this simplifying condition of commutation is expressed in terms of $\mathbf{t^*}$, for a switched (singular) system.

Another point is that in order to stress the specificity of (9), and the possible mutual independence of $t_k$, a specific (proper) *geometric representation* of the solutions of such singular state-space problems, *using a separate axis for each $t_k$*, should be developed.

**E**:  Some interesting geometric presentations can be found in the theory of "sliding-mode" (also some switching) systems, e.g. [18,19], but I agree that the exceptional analytical role of $t_k$ deserves more treatment, even from the positions of teaching the theory of such singular systems.

**P**:  This deepening into mathematics is boring! There are much simpler, feasible things that interest any physicist. First of all, the shifting nonlinearity relates to the topic of "kicked oscillators". If the instants at which the pulses of a force act on such an oscillator are dependent on the solution function $x(t)$, -- then $x(t)$ can be a chaotic function/process. Thus, in [20] the following, *per se* linear and non-singular, equation ($a, b, c$ -- constants)

$$\frac{d^2x}{dt^2} + a\frac{dx}{dt} + bx(t) = c\sin\omega t \qquad (24)$$

is considered with, however, the singular "forced" condition of the mirror-type reflection of $x(t)$ from the time axes at the zeros of $x(t)$, which causes $x(t)$ to always be of the same sign:

$$(dx/dt)(t_k^+) = -(dx/dt)(t_k^-), \qquad t_k: x(t_k) = 0. \qquad (25)$$

Equations (24,25) define *together* a *nonlinear* system with the relevant nonlinearity caused by the use in (25) of the zeros of the function to be found. Computer experiment has shown [20] that, for some ranges of the parameters $a$, $b$ and $c$, the sequential development of the process leads to a chaotic $x(t)$, which is a feature of a nonlinear system.

**E**: This is a nice example, but I am afraid that today such results are already not surprising.

**P**: Then let me pass on to a much more basic physics problem using your point of view. This problem even seems to me to be more interesting than all those preceding in our discussion. Let me start from the following question.
   *If we consider chaos to be a typical nonlinear feature of the relatively simple systems that we have in our laboratories and model on our computers, -- then why not try to see the chaotic movements of the molecules of a gas as an expression of a nonlinearity?*
   Thinking about the "shifting nonlinearity", I am going to consider "singular" collisions between hard balls; the colliding balls to be a simplified model of the dynamics of the molecules' ensemble. Of course, some "softer" interactions between the particles in an ensemble can lead to the statistical (thermodynamic) equilibrium and the chaoticity of movement. However, the model of an ideal gas is widely used in physics.

**E**: A simplified straightforward model should be useful here. However, it seems that you can formulate your thesis more decisively:
   *Since in any statistical equilibrium there are chaotic movements of the particles, and chaos may be obtained only in an NL system, any statistical ensemble that can be in an equilibrium state is nonlinear, i.e. is an NL system.*
   If so, -- could it be that the shifting nonlinearity is "as old as this world"?

**M**: I like your enthusiasm, but suggest you to make the nonlinearity of the problem be *directly* obvious. I am not quite sure that the statistical character of the problem cannot lead to the chaos without a nonlinear mechanism. Do not be surprised; the concept of "system" that you use is not so simple. In the case of liquid flow, you assumed that this concept can be applied, but since we still have no general solution of the nonlinear state equations, I can not be sure that every ensemble of interacting particles can/should be seen as a "system" in the state-space sense.
   Thus, you should be more concrete.

**P**: Consider the establishment of statistical equilibrium ("thermalization") in an ensemble of absolutely rigid small colliding balls, each of mass '*m*'. For many balls, a detailed description of the dynamics of the ensemble is very difficult even for computer simulation. However, it is sufficient to look at the equations.

The spatial position $r_i(t)$ of a ball number '*i*' is defined by its initial position and the $\delta(t)$-type forces that arise in the collisions of the ball with other balls at the instants $t_{i,j}$ that are roots of equations of the type $r_i(t) - r_j(t) = 0$, $j \neq i$. These forces may be written as $\boldsymbol{F}_{ij} = \boldsymbol{F}_o(\boldsymbol{v}_i(t_{ij}^-), \boldsymbol{v}_j(t_{ij}^-)) \delta(t-t_{ij})$, where $\boldsymbol{v}_i = d\boldsymbol{r}_i/dt$, i.e. $\boldsymbol{v}_i(t_{ij}^-)$ and $\boldsymbol{v}_j(t_{ij}^-)$ are the velocities of the balls just before the collision. Since the function $F_o(.,..)$ can be found from energy and impulse conservation laws, its very form is universal for the collisions, and let us focus only on the time-dependent factor $\delta(t-t_{ij})$ that includes the shift.

In view of the dynamic law

$$\frac{d\vec{v}_i}{dt} = \frac{1}{m} \sum_j \vec{F}_{ij} = \\
= \frac{1}{m} \sum_j \vec{F}_o(v_i(t_{ij}^-), v_j(t_{ij}^-)) \delta(t-t_{ij}), \\
j \neq i, \; \forall i, \tag{26}$$

$\{v_i(t)\}$, and thus $\{r_i(t)\}$, may be presented as some explicit functions of $\{t-t_{ij}\}$. However since $\{t_{ij}\}$ themselves are defined by the unknown $\{r_i(t)\}$ (our **x** here), (26) is a system of nonlinear equations in the sense of the shifting nonlinearity. Comparing (26) with (18) or (19), I see the $\{t_{ij}\}$ as some "**t*(x)**".

Thus, I can conclude that the process of the establishment of the statistical equilibrium of the velocities' distribution in the ensemble of the balls is an essentially nonlinear process. Believing that nonlinearity is the main point, I hope that now my position is more stable against the mathematical criticism.

**E**: The ensemble of the colliding balls now seems to me to be a kind of switched system. However, I also agree with the mathematical position that the limits for application of the concept of "system" have to be considered for such problems. Indeed, for a system, one can require performing the "input-output" test of linearity, etc.. I would feel myself better if there would be a sensor at a point, registering the impulses of the particles, this signal to be an "output" of the "system".

**M**: Both of you can see that my criticism is an encouraging one. The nonlinear equations support your philosophy, but since chaotic movements are met in very different statistical ensembles, you should at least also prove that *any* realistic interaction between colliding particles is nonlinear.

However, -- to what degree is the statistical nature of the problem with the ensemble of the balls important for the "nonlinearity approach"? If there are only two balls, then one can calculate the process starting from some given initial conditions, and then $t_{ij}$ appearing in $\delta(t-t_{ij})$ are *known*.

**P**:  This question is to the point, but this time I have to stay aside because, regrettably, I do not know how to measure the initial conditions even for *one* molecule [22].

**E**:  I see that for the microscopic particles it would be difficult to perform such a measurement, but in electrical circuits it is always possible to give/define the initial conditions.   In my terms, the mathematical comment about two balls concerns *independent small/simple subsystems*, and I was sure that we shall come to a relevant example.

I shall let you decide how to "translate" the following electrical-circuit example to the situations with the balls, but this example well shows how sensitive are the things to the complexity of a system.  You will see that a nonlinear singular system can become linear with a very simple decoupling.  There is at least a partial answer here to the question asked, since one sees that sometimes, in such systems, "to be linear" means "to be independent", and it is interesting that the term "independent" can be related only to one part of a system, in the sense of a "master-slave" relation between the parts.

In the following circuit (Fig. 6) $v_{inp}(t)$ is known, which makes the whole system "closed", and the reference-voltage of the comparator, $v_{ref}(t)$, is also known; it may even be given by a simple battery.

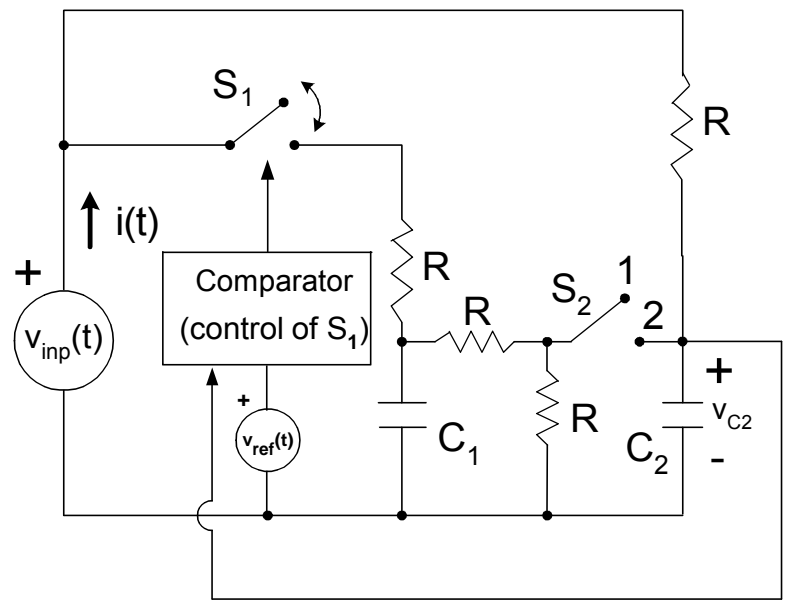

Fig. 6: A circuit illustrating the possibility of separating a nonlinear switched system into two linear subsystems.  The circuit becomes linear when switch $S_2$ is opened, i.e. non-conductive. Consider that the triggering input of the comparator has very large ("infinite") impedance i.e. takes a negligible current.  That is, the connection of point '2' to the comparator does not influence $v_{C2}(t)$.

Consider first switch $S_2$ to be in position 1.  Then $v_{C2}(t)$ is simply independently defined by the input voltage $v_{inp}(t)$ and we can consider $v_{C2}(t)$ to be known.  Thus, having two known functions/signals at the inputs of the comparator we can control switch $S_1$ to be operated at some prescribed (known) instances and thus see the system in whole as LTV. You may note that in this case the state equations are *split* in the sense that the equation for $x_2(t) \equiv v_{C2}(t)$ is independent from the other state-equation written for $x_1(t) \equiv v_{C1}(t)$. Equivalently, when $S_2$ is open the system is presented as two linear systems, one (with $C_2$) time-invariant, and the other (with $C_1$) time-variant. The first is the "master" and the second the "slave".

Thus, if $S_2$ is in position 1, then any function in the system, for instance the input current $i(t)$, depends on $v_{inp}(t)$ *linearly*; as a convolution of $v_{inp}(t)$ with some impulse response $h(t,\lambda)$. By doubling $v_{inp}(t)$ you can double $i(t)$, etc.. This scaling linearity does not fit any chaotic response, and, certainly, no chaos can be obtained in any such (any linear) system.

If, however, $S_2$ is in position 2, then there is no independent subcircuit for $C_2$, and the given circuit, or the system of the equations, has to be considered as a whole. Then, since the trigger input of the comparator depends on the unknown state variable $x_2(t) \equiv v_{C2}(t)$ by itself influenced by $x_1(t) \equiv v_{C1}(t)$, the control of $S_1$ is *not* known a priori, and the system is *nonlinear*. Then $i(t)$, -- like any other function being generated in the circuit, -- is *not* a linear response to $v_{inp}(t)$, and by doubling $v_{inp}(t)$ we shall *not* obtain, in general, a doubling of $i(t)$.

**P**: I agree with your proposition that in some such systems, to become linear means to become independent. In fact, already when you combined Fig. 1 with the state-space analysis, I noted that a dependence of the $f_{trig}$ on some $x_s$, i.e. the appearance of the nonlinearity can mean that the switched unit with the trigger ceases to be an independent subsystem.

**M**: This possibility is also seen via comparison of (5) with (4), if some relevant subsystems are considered. One just has to see [A] and [B] as composed of some smaller matrices. Perhaps this item can stress the advantage of use (5) instead of (12) as regards the structural aspect.

However, can the present circuit generate chaos for a constant input voltage, or respond chaotically to, say, a sinusoidal one?

**E**: I cannot immediately say whether or not chaos can be obtained in this circuit. Specialists in chaos study the conditions for chaos to be obtained. You can give this circuit to a student using PSpice or any other modeling. However, the general possibilities of electronic design are very rich and chaos has been observed in many electrical circuits. Even in some relatively simple power electronics circuits, *whose nonlinearity should now be obvious to you*, the conditions for chaos have been found. There are many such works, e.g., in the IEEE Transactions on Circuits and Systems, and see, e.g., the works cited in [11,21].

However, we speak too much about chaos. Nonlinearity is usually very important for the stability of the amplitude of some periodic oscillations in practical oscillators, which relates, of course, to the simple *limit cycles* that are the prototypes of the chaotic attractors.

Concluding my point, I would like to say that when joining you, I just intended to use the realization argument to help you to see the difference between linearity and nonlinearity of some modern systems. I hope that my reasoning was of some help, and for me, your physical arguments and mathematical comments were important. Last, but not the least, I am grateful to you for not being prejudiced in any sense. Everyone has his specialization, but for a scientific discussion it is not always good to "know what one needs" ... .

I hope that in the not too distant future, some basic concepts of singular electronics systems will take their place in general physics textbooks, much as the classical LCR circuits do today (see, e.g., [23]). A good textbook entitled "*System Theory for Physicists*" would be useful, and a good university-level laboratory named "Switching Nonlinearity", or "Singular Systems", which would combine the engineering, physics and mathematical outlooks, should be interesting to a serious student.

## 3. Conclusions

The shifting nonlinearity can be qualitatively formulated in terms of time functions, not necessarily characteristics; in the general case we deal here not with separate elements, but with some subsystems including also triggers, comparators and samplers. The point is to observe whether or not the triggers are dependent on the state variables to be found. This very general outlook on switched (singular) circuits and systems, developed here and in [2,3], should be heuristically useful. More standard facts about basic nonlinear circuits can be found, e.g., in the classical works [5,24,25].

In the most general terms, one should consider whether or not the *structure* of a "system" depends on the variables to be found, and the distinction between (5) and (12) is relevant for understanding what the "structure" of a system is. It is not trivial to see the vector field of a flow not only as the unknown vector-function to be found, but also as the "structure" of the liquid "system". It is also unusual to suppose that since the particles of an ensemble are moving chaotically, this ensemble should be a nonlinear system.

Circuits examples given here, and more in [2,3], may be found interesting for analyzing and modeling by students.

Besides the formal purpose to help one to see whether a singular system is of the type (4) or (5), i.e. linear (LTV) or nonlinear, the pedagogical position of the present work is also the stressing that circuit theory is not just a branch of applied mathematics or some numerical simulations. Quite similarly to the seeing this world via basic physics and mathematics, seeing it also via some basic system theory, in particular via the concept of "shifting nonlinearity", should be a part of one's general culture. There is even the hope here that the present work will be motivating, causing one to find new interesting points of view on singular systems, and new connections with other fields, say (perhaps), with differential-games theory where nonlinear differential equations also arise.

In any case, the given discussion allows one to better understand *what nonlinearity is* (*can be*). The opinion here is that the "**t***(**x**)-nonlinearity" is not less important, than, say, the nonlinearity introduced by a cubic characteristic, and, certainly, much more interesting. The heuristic/pedagogical aspect is here obvious.

## Acknowledgements

I am grateful to Professor Steven Shnider for some comments and advices regarding the focuses, made on initial version of the work. My understanding of switched (singular) circuits partly originated from my old research on fluorescent lamp circuits. I am thus indebted to Professor Menachem Krichevsky for my initial acquaintance with the lamp circuits, and to Professor Ben-Zion Kaplan in whose laboratory I worked on these circuits as a PhD student many years ago. Some important help by Peter Lambert is also appreciated on this occasion.

**References:**

[1]  I. M. Gel'fand, "*Lectures on Linear Algebra*", Interscience, New York, 1961.

[2]  E. Gluskin, "A point of view on the linearity and nonlinearity of switched systems", Proceedings of 2006 IEEE 24th Convention of Electrical and Electronics Engineers in Israel, (Eilat, Nov. 2006), pp. 110-114.  (The work is found in IEEE XPlore publication web-site.)

[3]  E. Gluskin, "A small theorem on the nonlinearity of switched systems", AEU - Int. Journal of Electronics and Communications.   In press, doi:10.1016/j.aeue.2007.04.006 .

[4]  L.A. Zadeh, Ch.A. Desoer, "*Linear System Theory*", McGraw-Hill, New York, 1963.

[5]  L.O. Chua,  "*Introduction to Nonlinear Network Theory*", McGraw Hill, New York, 1969.

[6] W.J. Rugh, "*Linear System Theory*", Prentice Hall, New York, 1980.

[7]  P.A. Fuhrmann, "*Linear Operators and Systems in Hilbert Space*", McGraw-Hill, New York 1981.

[8]  J. von Neumann, "*Mathematische Grunglagen der Quantenmechanik*", Berlin, Verlag von Julius Springer, 1932.

[9]  L.D. Landau and E.M. Lifshitz,  "*Fluid Mechanics*", Elsevier, Amsterdam, 2004.

[10] E. Gluskin, "On the complexity of a locally perturbed liquid flow", Physics Letters A, 183, 1993 (175-186).

[11]  P.K. Roy, A. Basuray and E. Gluskin, "On power supplies used for laboratory demonstration of chaotic electronic oscillators", American Journal of Physics, American Journal of Physics. Vol. 73, no. 11 (Nov. 2005), pp 1082-1085.

[12]  E. Gluskin, "A zerocrossing analysis", Progress in Nonlinear Science Proc. of the Int'l. Conf. dedicated to the 100th Anniversary of A.A. Andronov Mathematical Problems of Nonlinear Dynamics, Nizhny Novgorod, Russia, I, 2-6 July 2001 (241-250)  (In Web only abstract appears; the full text published in the Proceedings may be taken from the homepage of the author.)

[13]  E. Gluskin, "The fluorescent lamp circuit", IEEE Transactions on Circuits and Systems, Pt.I: Fundamental Theory and Applications vol. 46, no. 5 (May 1999), pp. 529-544. (This is a CAS "Exposition").

[14]  E. Gluskin, "The nonlinear theory of fluorescent lamp circuits", Int'l. J. of Electronics, vol. 63 (1987), pp. 687-705.  **DOI:** 10.1080/00207218708939175

[15]  E. Gluskin, "On the theory of an integral equation", Advances in Applied Mathematics, 15(3), 1994 (305-335)

[16] E. Gluskin, "The power consumed by a strongly nonlinear element with a hysteresis characteristic fed via a periodically driven LC circuit", J. of the Franklin Institute, 328(4), 1991 (369-377)

[17] E. Gluskin, F. V. Topalis, I. Kateri and N. Bisketzis,"The instantaneous light-intensity function of a fluorescent lamp", Physics Letters A, Volume 353, no. 5, 8 May 2006, pp. 355-363.

[18] V. Utkin, "*Sliding Modes in Control and Optimization*", Springer, 1992.

[19] D. Liberzon, "*Switching in Systems and Control*", Birkhauser, 2003.

[20] H. Isomaki, J. von Boehn and R. Raty, "Devil's attractors and chaos of a driven impact oscillator" *Phys. Lett. A*, 1985, **107**A(8), 343-346.

[21] O. Feely, "Nonlinear dynamics of discrete-time circuits: A survey", Int'l J. of Circuit Theory and Applications, vol. 35 (2007), pp. 515-531.

[22] E. Gluskin, "Nonlinear systems: between a law and a definition", Reports on Progress in Physics", vol. 60, no. 10, (Oct. 1997), 1063-1112

[23] R.P. Feynman, R.B. Leighton, M. Sands, "*The Feynman Lectures on Physics*", vol. 1, Addison-Wesley Publishing Company, 1970.

[24] L.O. Chua, Ch.A. Desoer, and E.S. Kuh, "*Linear and Nonlinear Circuits*", McGraw-Hill, New York, 1987 (see p. 85).

[25] L.O. Chua, "A computer-oriented sophomore course on nonlinear circuit analysis", IEEE Transactions on Education*,* vol. E-12, no. 3 (Sept. 1969), pp. 202-208.